\documentclass[11pt]{article}
\usepackage[a4paper,margin=28mm]{geometry}
\usepackage{amsmath,amssymb,amsthm,mathtools}
\usepackage{mathrsfs}
\usepackage{graphicx,booktabs,microtype}
\usepackage[colorlinks=true,linkcolor=blue,citecolor=blue,urlcolor=blue]{hyperref}
\usepackage[nameinlink,noabbrev]{cleveref}
\numberwithin{equation}{section}
\newtheorem{theorem}{Theorem}[section]
\newtheorem{proposition}[theorem]{Proposition}
\newtheorem{lemma}[theorem]{Lemma}
\newtheorem{corollary}[theorem]{Corollary}
\theoremstyle{definition}
\newtheorem{definition}[theorem]{Definition}
\theoremstyle{remark}
\newtheorem{remark}[theorem]{Remark}
\newcommand{\R}{\mathbb R}
\newcommand{\C}{\mathbb C}
\newcommand{\U}{\mathbb U}
\newcommand{\Sp}{\operatorname{Sp}}
\newcommand{\diag}{\operatorname{diag}}
\newcommand{\tr}{\operatorname{tr}}

\newcommand{\ip}[2]{\langle #1,#2\rangle}
\newcommand{\dd}{\,dt}

\title{Precise Analytic Boundaries and Positive Symplectic Paths\\
in the Stability Diagrams of Elliptic Relative Equilibria}
\author{Jian Wang \thanks{Partially supported by NSFC (Nos. {12361141812, 12071231}) and the Fundamental Research Funds for the Central University (No. 63243062). E-mail: wangjian@nankai.edu.cn} \quad and \quad
	Qinglong Zhou \thanks{Partially supported by NSFC (No.12171426), the Natural Science Foundation of Zhejiang Province (No. LY19A010020) and the Fundamental Research Funds for the Central Universities (No. 226-2024-00136).
		E-mail: zhouqinglong@zju.edu.cn. }\\
	$^{1}$ School of Mathematical Sciences and LPMC,\\ Nankai University, Tianjin 300071, China\\
	$^{2}$ School of Mathematical Science,\\Zhejiang University, Hangzhou 310058, Zhejiang, China\\
}
\date{}
\begin{document}
\maketitle
\begin{abstract}
We investigate the dependence of the linear stability of elliptic relative
equilibria on the mass parameter and eccentricity. Using a periodic matrix
Riccati decomposition, we derive explicit analytic curves separating
hyperbolic--hyperbolic and complex--saddle spectra. 
The construction applies to the Lagrangian,
Euler--Robe, and $(2+2)$-body families. A Hermitian Riccati factorization
also yields an eccentricity-independent sufficient condition for positivity
of the associated operators under every unit-circle boundary condition,
and hence for hyperbolicity of the monodromy matrix. 
Finally, we construct an explicit lower bound for the second
antiperiodic degeneracy curve by a finite trigonometric test function.
This bound agrees with the value and tangent of the degeneracy curve at
the circular limit. Together, these results provide analytic descriptions
and estimates for distinct spectral boundaries in the stability diagrams.
\end{abstract}

\section{Introduction}
The collinear motions of Euler and the equilateral motions of Lagrange
are classical examples of homographic solutions of the Newtonian three-body
problem. Their configurations retain their shape while their size and
orientation evolve according to a Keplerian motion. More generally, a central
configuration generates an elliptic relative equilibrium. In the circular
Lagrangian problem, the Gascheau--Routh criterion expresses stability in
terms of the masses. For nonzero eccentricity, the variational equation has
periodic coefficients, and the stability diagram depends on both a mass
parameter and the eccentricity \cite{Roberts2002,MS2005}.

The same framework includes the planar restricted problem. In the limit
where one mass vanishes, the collinear configurations yield the three
collinear libration points, while the two orientations of the equilateral
configuration yield the triangular points. For elliptic primary motion these
configurations are fixed in rotating-pulsating coordinates. Their essential
variational equations agree with the corresponding Euler or Lagrangian
families after the appropriate parameter identification. The parameters of
the two families must nevertheless be distinguished: the Lagrangian mass
parameter ranges over $[0,9]$, whereas the physical Euler parameter ranges
over $[0,7]$ \cite{HLS2014,ZL2017}.
The linearized system of Robe's restricted three-body problem coincides
with the Euler essential system, but its mass parameter ranges over
$[-1,0]$ \cite{ZZ2017}. Accordingly, we sometimes consider the extended
parameter interval $[-1,7]$ when studying the linear stability of the
Euler family, thereby including the Robe problem.

Geometric reductions make these stability problems accessible to analysis.
Roberts \cite{Roberts2002} studied the essential four-dimensional
Lagrangian system and its stability regions. The symplectic coordinates of
Meyer and Schmidt \cite{MS2005} separate the center-of-mass, Keplerian,
and essential parts of the twelve-dimensional planar three-body variational
equation. The nontrivial Floquet multipliers belong to the essential part.
Mart\'{i}nez, Sam\`a, and Sim\'o investigated these periodic systems by normal
forms near the circular limit and by analysis of the singular collision
limit \cite{MSS2006a,MSS2006b}. In rotating-pulsating coordinates, with
true anomaly as the independent variable, their coefficients contain the
common factor $(1+e\cos t)^{-1}$. This structure also underlies Tschauner's
finite trigonometric construction for the matrix Riccati equation
\cite{Tschauner1973}.

A complementary approach uses symplectic and variational indices.
For a normalized symplectic path $\gamma$ and $\omega\in\U$, the
Maslov-type index pair $(i_\omega,\nu_\omega)$ records intersections with
the $\omega$-degeneracy set and the dimension of
$\ker(\gamma(2\pi)-\omega I)$. Splitting numbers and iteration formulas
relate these indices to the unit-circle spectrum \cite{Long2002,RS1993}.
For the second-order systems considered here, the Morse--Maslov
correspondence identifies the antiperiodic Morse index with $i_{-1}$.
Thus the parameter curves on which $-1$ is a Floquet multiplier can be
studied through self-adjoint boundary-value problems.

Hu and Sun developed symmetry-adapted index methods and applied the Morse
index to elliptic Lagrangian solutions \cite{HS2009,HS2010}.
Hu, Long, and Sun \cite{HLS2014} described the Lagrangian family throughout
the mass--eccentricity rectangle, including its two antiperiodic degeneracy
curves and the boundary of the hyperbolic region. Collision-index methods
address the singular limit $e\to1$ \cite{HO2016}, while Zhou and Long
\cite{ZL2017} established the indexed degeneracy curves for elliptic Euler
solutions. At fixed eccentricity, the antiperiodic Morse index decreases
with the Lagrangian parameter and increases with the Euler parameter;
the jump at a degeneracy is determined by its nullity. These orientations
are properties of the respective normalized families and cannot be
transferred merely by identifying their parameter symbols.

Ordering the degeneracy curves is distinct from proving their strict
separation. A simple kernel in each reflection sector does not by itself
exclude a coincidence of the two sector parameters. The recent preprint
of Hu, Wu, and Zhu \cite{HWZ2026} studies this issue for both Euler and
Lagrangian solutions. Its half-integer Fourier formulation isolates a
common infinite tail and a different lowest-frequency block in the two
reflection sectors. This perspective also explains the usefulness of
low-frequency antiperiodic test functions. In the present paper, such a
test function provides an explicit estimate for the location of the second
Lagrangian curve; the estimate is complementary to the qualitative
separation problem.

The spectral diagram therefore involves several different questions:
where reciprocal multiplier pairs collide, in which direction a
unit-circle degeneracy is crossed, and where an individual degeneracy
curve lies. An algebraic Riccati decomposition, a parameter crossing form,
and a variational test function address these questions, respectively.

The purpose of this paper is to connect three complementary descriptions of
the stability diagram. First, we express the Riccati discriminant explicitly
and relate it to the spectral discriminant of the monodromy matrix. This
relation separates the algebraic collision condition from the additional
conditions that make the collision an HH--CS boundary. Second, for the
Lagrangian family we derive a positive parameter crossing form on each
unit-circle eigenspace. Third, we use the resulting index interpretation to
obtain an explicit estimate for the right-hand, or second, antiperiodic
degeneracy curve. The last result is variational and does not require the
explicit solution of a Floquet equation.

Throughout, $\U=\{z\in\C:|z|=1\}$. A real symplectic $4\times4$ matrix is
called elliptic--elliptic (EE) when both reciprocal pairs lie in
$\U\setminus\R$, elliptic--hyperbolic (EH) when one pair lies there and the
other lies in $\R\setminus\{0,\pm1\}$, hyperbolic--hyperbolic (HH) when all
multipliers lie in $\R\setminus\{0,\pm1\}$, and complex--saddle (CS) when
all multipliers lie outside $\U\cup\R$. Multiplicities are counted
algebraically. The HH and CS regions are both hyperbolic. Thus a boundary
between them is distinct from the boundary of the full hyperbolic region.

For reference, the principal conclusions are the following.
\begin{theorem}\label{thm:intro-boundaries}
The local boundaries separating HH and CS spectra are given by the
following curves.
\begin{enumerate}
\item \textup{Lagrangian solutions.}
For $0<\beta_L<9$ and $0\le e<1$, the locus is
\[
\beta_L=5+e^2\pm2\sqrt{2(1-e^2)(2+e^2)}.
\]
\item \textup{Euler--Robe family.}
For $-1<\beta_E\le7$, the locus lies in the
Robe range $-1<\beta_E\le-1/9$ and is given by
\[
e^2=\sqrt{-2\beta_E(\beta_E+1)(10\beta_E^2+27\beta_E+18)}
-\frac12(\beta_E+1)(\beta_E+4).
\]
Only values with $0\le e^2<1$ are included. Its closure meets $e=0$
at $\beta_E=-1,-1/9$ and $e=1$ at
$\beta_E=(\sqrt{73}-17)/18$.
In the physical Euler range $0\le\beta_E\le7$, there are no CS
spectra and hence no HH--CS boundary.
\item \textup{Elliptic relative equilibria of the $(2+2)$-body problem.}
Let $\beta_F=27m(1-m)\in[0,27/4]$ and $s=\sqrt{9-\beta_F}$.
For the convex family, the locus is
\[
e^2=\frac{4|5s-9|\sqrt{7s(3-s)(s+1)(25s+3)^3}
-(5s-9)^2(75s^2+112s-3)}{24(11s+9)^2},
\]
restricted to
\[
\frac{16(182-37\sqrt{21})}{625}\le\beta_F\le\frac{27}{4},
\qquad \beta_F\ne\frac{144}{25},\qquad 0\le e^2<1.
\]
Its closure meets $e=0$ at the left endpoint and at $\beta_F=144/25$,
and $e=1$ at $\beta_F=1656/625$.
For the nonconvex family, there are no CS
spectra and hence no HH--CS boundary.
\end{enumerate}
The points with $e=1$ and the excluded parameter values are limiting
points only.
\end{theorem}
The construction uses a periodic Riccati decomposition with
$c_1\ne c_2$, $(c_1+c_2-3)^2\ne4c_1c_2$, and $|eq|<1$, where
$q$ is defined in \eqref{eq:nondeg}. Its discriminant $\Delta$ and the
spectral discriminant $\mathscr D$ satisfy
$\mathscr D=\Delta\kappa^2$, with a real analytic trace factor $\kappa$.
The boundary assertion in \cref{thm:intro-boundaries} applies at points
of $\Delta=0$ where $d\Delta\ne0$, $\kappa\ne0$, and the common
reduced trace satisfies $|\tau_0|>2$. Locally, the $\Delta>0$ side is
HH and the $\Delta<0$ side is CS. The coefficients, discriminants, and
parameter ranges for each family are given in \cref{sec:analytic};
the spectral implication is proved in \cref{thm:spectral}.

The Lagrangian HH--CS boundary formula was first obtained by Tschauner
\cite{Tschauner1973} for the triangular points of the elliptic restricted
three-body problem. Under the parameter identification $\beta_L=9g$,
his formula gives the Lagrangian boundary displayed in
\cref{thm:intro-boundaries}.

For the Lagrangian family, let
$\gamma_{\beta_L,e}(t)\in\Sp(4)$ denote the fundamental matrix of the
essential variational equation in rotating-pulsating coordinates,
normalized by $\gamma_{\beta_L,e}(0)=I_4$. More precisely,
\[
\gamma_{\beta_L,e}'=J_4
\begin{pmatrix}I_2&-J_2\\J_2&I_2-\dfrac{D_{\beta_L}}{1+e\cos t}\end{pmatrix}
\gamma_{\beta_L,e},\qquad
D_{\beta_L}=\frac12\diag(3+\sqrt{9-\beta_L},3-\sqrt{9-\beta_L}),
\]
where $J_2=\left(\begin{smallmatrix}0&-1\\1&0\end{smallmatrix}\right)$
and $J_4=\left(\begin{smallmatrix}0&-I_2\\I_2&0\end{smallmatrix}\right)$.
The eigenvalues of $\gamma_{\beta_L,e}(2\pi)$ are the nontrivial Floquet
multipliers governing linear stability.
The local boundary description can be combined
with monotonicity of the Morse indices with respect to the mass parameter to obtain a larger hyperbolic region.
Indeed, once all unit-circle boundary-value operators are positive at a CS
parameter, their normalized monotonicity excludes unit-circle multipliers
at every larger parameter.

\begin{corollary}\label{cor:right-hyperbolic}
For any $0<e<1$,if
\[
e^2+5-2\sqrt{2(1-e^2)(2+e^2)}<\beta_L\le9,
\]
then $\gamma_{\beta_L,e}(2\pi)$ is hyperbolic, and the corresponding
elliptic Lagrangian solution is linearly unstable. In particular, this
holds for $6<\beta_L\le9$. The same conclusions hold at $e=0$ without
the additional spectral hypotheses.
\end{corollary}
\begin{proof}
Write $\beta_-(e)=e^2+5-2\sqrt{2(1-e^2)(2+e^2)}$ and fix
$\beta_L>\beta_-(e)$. By \cref{thm:intro-boundaries}, there is a CS
parameter $\beta_0\in(\beta_-(e),\beta_L)$. Thus
$\nu_\omega(A(\beta_0,e))=0$ for every $\omega\in\U$.
The Morse index is locally constant as the boundary multiplier varies
without degeneracy. Since $\U$ is connected, it follows that
\[
\phi_\omega(A(\beta_0,e))=\phi_1(A(\beta_0,e))=0
\qquad(\omega\in\U),
\]
where the last equality is the known periodic index property of the
Lagrangian family \cite{HLS2014}. Hence $A(\beta_0,e)$ is strictly
positive for every unit-circle boundary condition. The strict form
monotonicity
\[
\partial_\beta\left(\frac{A(\beta,e)}{\sqrt{9-\beta}}\right)
=\frac{K_e}{2(9-\beta)^{3/2}}>0
\]
proved in \cref{prop:monotone} implies that $A(\beta_L,e)>0$ for
$\beta_0<\beta_L<9$. At $\beta_L=9$, positivity follows from
\cref{lem:K}. Therefore no unit-circle multiplier can occur at any of
these parameters, proving hyperbolicity and linear instability.
Since $\beta_-(e)\le5+e^2<6$, the stated sufficient condition follows.
At $e=0$, the characteristic exponents satisfy
$\lambda^4+\lambda^2+\beta_L/4=0$. For $\beta_L>1=\beta_-(0)$,
none is purely imaginary, so the monodromy is hyperbolic as well.
\end{proof}

The positivity mechanism underlying Corollary~\ref{cor:right-hyperbolic}
extends beyond the Lagrangian family to systems with a general constant
diagonal matrix $D$. The following theorem gives an eccentricity-independent
sufficient condition for positivity under every unit-circle boundary
condition, and hence for hyperbolicity of the associated monodromy matrix.
For the Lagrangian family, $c_1+c_2=3$ and
$(c_2-c_1)^2=9-\beta_L$, so its condition reduces to $\beta_L>6$,
recovering the uniform sufficient condition in Corollary~\ref{cor:right-hyperbolic}.
Thus it extends that conclusion to the general diagonal family, whereas
Corollary~\ref{cor:right-hyperbolic} gives the sharper eccentricity-dependent
bound for the Lagrangian family.
\begin{theorem}\label{thm:intro-uniform-positive}
Let $D=\diag(c_1,c_2)$, $c=c_1+c_2>2$, and $d=c_2-c_1$.
Suppose that
\begin{equation}\label{eq:uniform-positive-region}
|d|<
\begin{cases}
2\sqrt7\,(c-2),&2<c<16/7,\\
\sqrt c,&c\ge16/7.
\end{cases}
\end{equation}
Then, for every $e\in[0,1)$, the operator
\[
A_D(e)=-\frac{d^2}{dt^2}I_2-I_2+
\frac{R(t)DR(t)^T}{1+e\cos t}
=-\frac{d^2}{dt^2}I_2-I_2+
\frac{cI_2-dS(t)}{2(1+e\cos t)},
\]
is strictly positive on $\overline D(\omega,2\pi)$ for any $\omega\in\U$.
\end{theorem}
The proof is given in \cref{subsec:uniform-positive}. We next return to
the Lagrangian family and its parameter crossing forms.

Fix $e\in[0,1)$, put $R(t)=e^{tJ_2}$ and $H=\diag(-I_2,I_2)$, and define
\[
\xi_{\beta,e}(t)=H\diag(R(t),R(t))\gamma_{\beta,e}(t)H,
\qquad M_e(\beta)=\xi_{\beta,e}(2\pi).
\]
Thus $\xi_{\beta,e}(0)=I_4$ and
$M_e(\beta)=H\gamma_{\beta,e}(2\pi)H$, so the two monodromy matrices
have the same spectrum. The matrix $H$ is antisymplectic and reverses
the sign convention for the parameter generator.
\begin{theorem}\label{thm:intro-crossing}
For $0<\beta<9$, define the real symmetric matrix
\[
\mathcal C_{\beta,e}
=-J_4\bigl(\partial_\beta M_e(\beta)\bigr)M_e(\beta)^{-1}.
\]
Then, for every $\omega\in\U$,
\[
\eta^*\mathcal C_{\beta,e}\eta>0
\qquad\text{for all }0\ne\eta\in\ker(M_e(\beta)-\omega I_4).
\]
In particular, every unit-circle crossing form of the parameter path
$\beta\mapsto M_e(\beta)$ is positive definite on its eigenspace.
\end{theorem}
The proof is given in \cref{thm:crossing}. The conclusion concerns the
unit-circle eigenspaces and does not assert positivity of
$\mathcal C_{\beta,e}$ on the entire phase space.

\begin{theorem}\label{thm:intro-estimate}
Let $\beta_m(e)$ denote the larger of the two antiperiodic degeneracy
parameters of the elliptic Lagrangian family at eccentricity $e$.
For $0\le e<1$, define
\begin{equation}\label{eq:b}
b(e)=9-\frac1{16}\left[
\sqrt{e^2+144+2\bigl(1+\sqrt{1-e^2}\bigr)
\bigl(12-15\sqrt{1-e^2}\bigr)}-e\right]^2.
\end{equation}
Then
\[
\beta_m(e)>b(e)\qquad\text{for every }0<e<1.
\]
Moreover,
\[
b(0)=\beta_m(0)=\frac34,\qquad
b'(0)=\beta_m'(0)=\frac{\sqrt{33}}4,\qquad
\lim_{e\to1^-}b(e)=0,
\]
where the derivatives at $e=0$ are right derivatives.
\end{theorem}
Thus \eqref{eq:b} gives an explicit strict lower bound for the second
antiperiodic degeneracy curve, with the same value and tangent at the
circular limit. The proof is given in \cref{thm:trial,cor:second}.

The paper is organized as follows. \Cref{sec:framework} introduces the
essential Hamiltonian systems and the associated self-adjoint operators,
fixes the boundary conditions and index conventions, and derives the matrix
Riccati equations governing invariant graphs. \Cref{sec:analytic} proves
\cref{thm:intro-boundaries}: the periodic graph decomposition yields the
factorization of the monodromy characteristic polynomial, and
\cref{thm:spectral} supplies the spectral criterion for an HH--CS boundary;
\cref{subsec:uniform-positive} proves \cref{thm:intro-uniform-positive}
by a Hermitian Riccati factorization and convex interpolation. The
subsequent subsections compute the boundary equations and parameter
ranges for the Lagrangian, Euler--Robe, and $(2+2)$-body families, followed
by an application to an even polygon block. \Cref{sec:positive} establishes
the positivity of a scalar comparison operator and the resulting parameter
monotonicity, then proves \cref{thm:intro-crossing} through the monodromy
crossing-form identity in \cref{thm:crossing}. Finally,
\cref{sec:estimate} proves \cref{thm:intro-estimate}: the explicit
antiperiodic trial function in \cref{thm:trial} gives a negative direction,
and \cref{cor:second} converts this into the strict bound for the second
degeneracy curve and verifies its endpoint and tangency properties.
Constant-denominator comparison estimates complete that section.
\Cref{app:riccati} contains the detailed scalar reduction and trigonometric
coefficient calculation underlying the explicit Riccati solutions.

\section{Hamiltonian systems and boundary-value operators}\label{sec:framework}
Let $I_n$ be the $n\times n$ identity matrix and 
$J_{2n}=\left(\begin{matrix}0&-I_n\\I_n&0\end{matrix}\right)$
be the standard symplectic matrix
for any $n\in\mathbb{N}$, omitting the subscript when no confusion arises.
Put
\[
R(t)=e^{tJ_2},\qquad N=\diag(1,-1).
\]
Let $0\le e<1$ and $D=\diag(c_1,c_2)$, where $c_1,c_2$
are real constants. We consider
\begin{equation}\label{eq:ham}
\gamma'=J_4B_D(t)\gamma,\qquad \gamma(0)=I_4,
\end{equation}
where
\begin{equation}\label{eq:B_D}
B_D(t)=\begin{pmatrix}I_2&-J_2\\J_2&I_2-D/(1+e\cos t)\end{pmatrix}.
\end{equation}
Writing a solution as $(p,q)^T$ gives $p=q'+J_2q$ and
\begin{equation}\label{eq:second}
q''+2J_2q'=Dq/(1+e\cos t).
\end{equation}
For $x=Rq$, this becomes $A_D(e)x=0$, where
\begin{equation}\label{eq:AD}
A_D(e)=-\frac{d^2}{dt^2}I_2-I_2+\frac{RDR^T}{1+e\cos t}.
\end{equation}
They are precisely the operators in \cite[Eq. (2.42)]{HLS2014}
 and \cite[Eq. (2.29)]{ZL2017} for some proper matrices $D$.
For $\omega\in\U$, its self-adjoint realization has domain
\[
\overline D(\omega,2\pi)=
\{x\in W^{2,2}([0,2\pi],\C^2):
x(2\pi)=\omega x(0),\ x'(2\pi)=\omega x'(0)\}.
\]
Its form domain consists of $W^{1,2}$ functions with the first of these
boundary conditions. The $\omega$-Morse index $\phi_\omega$ counts negative
eigenvalues with multiplicity, and $\nu_\omega$ denotes the dimension of the
kernel. The periodic change of coordinates implies
\begin{equation}\label{eq:nullity}
\nu_\omega(A_D(e))=\dim_\C\ker(\gamma(2\pi)-\omega I_4).
\end{equation}

For example, if we consider the elliptic Lagrangian solutions,
we have
\begin{equation}\label{eq:lagD}
D_\beta=\frac12\diag(3+\sqrt{9-\beta},3-\sqrt{9-\beta}),
\end{equation}
where $\beta=\frac{27(m_1m_2+m_2m_3+m_3m_1)}{(m_1+m_2+m_3)^2}\in[0,9]$.
Since $RNR^T=S(t)$, with
\[
S(t)=\begin{pmatrix}\cos2t&\sin2t\\\sin2t&-\cos2t\end{pmatrix},
\]
we obtain
\begin{equation}\label{eq:A}
A(\beta,e)=-\frac{d^2}{dt^2}I_2-I_2+
\frac{3I_2+\sqrt{9-\beta}S(t)}{2(1+e\cos t)}.
\end{equation}
Note that the endpoint $e=1$ is singular and is used only for limits of algebraic
curves, not as a regular periodic boundary-value problem.
\medskip

To obtain a decomposition of \eqref{eq:ham}, consider a time-dependent
invariant graph $p=Q(t)q$, where $Q(t)$ is a $2\times2$ matrix.
The Hamiltonian equations are
\[
p'=-J_2p-\left(I_2-\frac{D}{1+e\cos t}\right)q,
\qquad q'=p-J_2q.
\]
On the graph, $q'=(Q-J_2)q$, and differentiation of $p=Qq$ gives
\[
p'=Q'q+Qq'=\bigl(Q'+Q(Q-J_2)\bigr)q.
\]
Comparing this expression with the equation for $p'$ shows that the graph
is invariant precisely when $Q$ satisfies the matrix Riccati equation
\begin{equation}\label{eq:riccati}
Q'=-Q^2+QJ_2-J_2Q-I_2+\frac{D}{1+e\cos t}.
\end{equation}
With $P=Q-J_2$, the equation on the graph is $q'=Pq$, and
\eqref{eq:riccati} is equivalently
\begin{equation}\label{eq:P-riccati}
P'+P^2=-2J_2P+\frac{D}{1+e\cos t}.
\end{equation}
The latter also follows directly by substituting $q'=Pq$ into
\eqref{eq:second}. No symmetry condition on $Q$ or $P$ is imposed.

If $Q_+$ and $Q_-$ are two solutions with $Q_+-Q_-$ invertible, their
graphs are complementary. The transformation
\[
\begin{pmatrix}p\\q\end{pmatrix}
=T(t)\begin{pmatrix}\xi_+\\\xi_-\end{pmatrix},
\qquad T(t)=\begin{pmatrix}Q_+(t)&Q_-(t)\\I_2&I_2\end{pmatrix},
\]
then yields two independent systems
\[
\xi_\pm'=P_\pm\xi_\pm,\qquad P_\pm=Q_\pm-J_2.
\]
When the graphs are smooth and $2\pi$-periodic, this decomposition also
relates their monodromy matrices to that of \eqref{eq:ham}.

\section{Analytic collision curves and hyperbolic spectral types}\label{sec:analytic}
\subsection{Periodic Riccati decomposition}
We now construct two explicit solutions $Q_\pm$ of \eqref{eq:riccati}
and determine when they give the periodic decomposition described above.
Following the construction of Tschauner \cite{Tschauner1973}, we obtain
particular solutions for a general constant diagonal matrix $D$ under
the nondegeneracy assumptions below. For the reader's convenience, the
details of the construction, including the scalar reduction and
coefficient comparison, are presented in \cref{app:riccati}.

Assume
\begin{equation}\label{eq:nondeg}
d:=c_2-c_1\ne0,\qquad
F:=(c_1+c_2-3)^2-4c_1c_2\ne0,
\qquad q=\frac{3(c_1+c_2-3)}F.
\end{equation}
Define
\begin{align}
p_1(t)&=(1+e\cos t)(1+eq\cos t),\label{eq:p1}\\
p_2(t)&=\frac d4+\frac{deq}{3}\cos t
-\frac{3e^2[1-(c_1+c_2-1)q]}{4d}\cos2t.\label{eq:p2}
\end{align}
The discriminant is
\begin{align}
\Delta={}&d^2-8(c_1+c_2-2)\nonumber\\
&+2e^2\left[1-\frac{7(c_1+c_2)-15}{3}q
+\frac43(c_1+c_2-3)q^2\right]\nonumber\\
&+\frac{9e^4}{d^2}[1-(c_1+c_2-1)q]^2.\label{eq:Delta}
\end{align}

\begin{proposition}\label{prop:riccati}
Under \eqref{eq:nondeg}, choose $c\in\C$ with $c^2=\Delta$. On each
interval where $p_1\ne0$, the matrices
\begin{equation}\label{eq:Q}
Q_\pm=\frac1{p_1}
\begin{pmatrix}
(p_1'-p_2')/2&p_2\mp c/4\\
p_2\pm c/4&(p_1'+p_2')/2
\end{pmatrix}
\end{equation}
solve \eqref{eq:riccati}. If $|eq|<1$ and $\Delta\ne0$, they give a
smooth periodic decomposition over $\R$ for $\Delta>0$ and over $\C$
for $\Delta<0$. Moreover,
\begin{equation}\label{eq:detT}
Q_+-Q_-=\frac{c}{2p_1}J_2,
\qquad \det T=\frac{\Delta}{4p_1^2}.
\end{equation}
\end{proposition}
\begin{proof}
Differentiating \eqref{eq:p1}--\eqref{eq:p2} gives
\begin{align}
p_2''+4p_2&=\frac{d}{1+e\cos t} p_1,\label{eq:scalar1}\\
p_1p_1''-\frac12(p_1')^2+\left(2-\frac{c_1+c_2}{1+e\cos t}\right)p_1^2
+2p_2^2+\frac12(p_2')^2&=\frac\Delta8.\label{eq:scalar2}
\end{align}
These identities imply \eqref{eq:riccati} by substitution in \eqref{eq:Q}.
The determinant follows from the displayed difference. Finally, $(1+e\cos t)>0$
and $|eq|<1$ imply $p_1>0$ on $\R$.
\end{proof}

\subsection{From the Riccati discriminant to a spectral boundary}
The algebraic equation $\Delta=0$ identifies a collision of the explicit
graphs. To determine its spectral meaning, retain the traces of the reduced
systems. Let $\Phi_\pm'=P_\pm\Phi_\pm$, $\Phi_\pm(0)=I_2$, and set
$\tau_\pm=\tr\Phi_\pm(2\pi)$. Since
$\tr P_\pm=p_1'/p_1$, Liouville's formula gives
$\det\Phi_\pm(2\pi)=1$. Periodicity of $T$ now yields
\begin{equation}\label{eq:char}
\det(\lambda I_4-\gamma(2\pi))
=(\lambda^2-\tau_+\lambda+1)(\lambda^2-\tau_-\lambda+1).
\end{equation}
For $\Delta>0$, both traces are real. For $\Delta<0$, choose
$c=i\sqrt{-\Delta}$; then $\tau_- =\overline{\tau_+}$.

\begin{theorem}\label{thm:spectral}
Suppose \eqref{eq:nondeg} and $|eq|<1$ hold in a real analytic parameter
neighborhood. There is a real analytic function $\kappa$ such that the
spectral discriminant satisfies
\begin{equation}\label{eq:spectraldiscriminant}
\mathscr D:=(\tr M)^2-4\left(\frac{(\tr M)^2-\tr(M^2)}2-2\right)
=\Delta\kappa^2,\qquad M=\gamma(2\pi).
\end{equation}
Consequently:
\begin{enumerate}
\item If $\Delta>0$, the spectrum cannot be CS. If it is hyperbolic, it is HH.
\item If $\Delta<0$ and $\kappa\ne0$, the spectrum is CS.
\item At a point where $\Delta=0$, write $\tau_0$ for the common limiting
trace. If $d\Delta\ne0$, $\kappa\ne0$, and $|\tau_0|>2$, then
$\Delta=0$ is an exact local HH--CS boundary: its positive side is HH and
its negative side is CS.
\end{enumerate}
\end{theorem}
\begin{proof}
Regard the right-hand side of \eqref{eq:Q} as an analytic expression in an
independent variable $c$, and let $\tau(c)$ be the resulting monodromy trace.
Its odd difference has the form
\[
\tau(c)-\tau(-c)=c\,k(c^2),
\]
where $k$ is analytic, with real coefficients for real parameters. Put
$\kappa=k(\Delta)$. For $\Delta\ne0$, \eqref{eq:char} gives
$\mathscr D=(\tau_+-\tau_-)^2=\Delta\kappa^2$. Analytic continuation
establishes the identity also at $\Delta=0$.
If the two traces are real, each reciprocal pair is either real or on the
unit circle. If they are nonreal, no root of either quadratic can be real
or have modulus one, because either condition would make
$\lambda+\lambda^{-1}$ real. These observations prove the first two claims.
At a point in the third claim, the limiting polynomial is
$(\lambda^2-\tau_0\lambda+1)^2$, with all roots off the unit circle.
Hyperbolicity is open. The signs of \eqref{eq:spectraldiscriminant}, together
with the regularity of the level set, then give the asserted local boundary.
\end{proof}

\begin{remark}\label{rem:scope}
The conditions $\kappa\ne0$ and $|\tau_0|>2$ distinguish an HH--CS
interface from an elliptic collision or a further trace resonance. Thus the
explicit equation $\Delta=0$ alone is not a classification of the entire
stability diagram. The theorem gives an exact boundary statement wherever
its stated spectral conditions hold, without identifying the boundary of
all hyperbolic parameters with this algebraic curve.
\end{remark}

\subsection{Uniform positivity for the diagonal family}\label{subsec:uniform-positive}
The explicit Riccati solutions also yield a positive factorization when
their discriminant is nonpositive. This gives the sufficient condition
stated in the Introduction.

\begin{proof}[Proof of \cref{thm:intro-uniform-positive}]
First let $c\ge16/7$ and $d=\sqrt c$. Then
\[
F=9-5c\ne0,\qquad q=-\frac{3(c-3)}{5c-9},\qquad |q|\le\frac{15}{17}<1,
\]
and \eqref{eq:Delta} simplifies to
\begin{equation}\label{eq:uniform-boundary-delta}
\Delta=-(1-e^2)\left[7c-16+
\frac{9c(3c-7)^2}{(5c-9)^2}e^2\right]\le0.
\end{equation}
Taking $\sqrt\Delta=i\sqrt{-\Delta}$ in \eqref{eq:Q} gives a smooth,
$2\pi$-periodic Hermitian solution $Q$ of \eqref{eq:riccati}.
This remains valid at $\Delta=0$ by direct substitution.
The Hermitian matrix $M=RQR^T$ consequently satisfies
\[
M'+M^2=-I_2+\frac{RDR^T}{1+e\cos t}.
\]
Integration by parts, with the boundary term vanishing because
$M(2\pi)=M(0)$ and $|\omega|=1$, gives
\begin{equation}\label{eq:uniform-factorization}
\langle A_D(e)x,x\rangle
=\int_0^{2\pi}|x'-Mx|^2\dd\ge0.
\end{equation}
The same argument applies to $d=-\sqrt c$, since $q$ and $\Delta$
depend only on $d^2$ and \eqref{eq:Q} is Hermitian for either sign of $d$.

For the isotropic matrix $(c/2)I_2$, the scalar ground-state identity gives
\begin{align*}
\langle A_{(c/2)I_2}(e)x,x\rangle
={}&\int_0^{2\pi}(1+e\cos t)^2
\left|\left(\frac{x}{1+e\cos t}\right)'\right|^2\dd\\
&+\frac{c-2}{2}\int_0^{2\pi}\frac{|x|^2}{1+e\cos t}\dd.
\end{align*}
Since $A_D(e)$ is affine in $D$, interpolation between the isotropic
matrix and the boundary matrix with $d=\pm\sqrt c$ yields, for
$|d|<\sqrt c$,
\[
\langle A_D(e)x,x\rangle\ge
\left(1-\frac{|d|}{\sqrt c}\right)
\frac{c-2}{2(1+e)}\|x\|_{L^2}^2>0\qquad(x\ne0).
\]

Now let $2<c<16/7$ and $|d|<2\sqrt7(c-2)$.
The case $c=2$, $d=0$ corresponds to $D=I_2$ and is nonnegative by
the same scalar identity. Convex combinations of this matrix with the
boundary matrices $(c,d)=(16/7,\pm4/\sqrt7)$ therefore give
nonnegative operators at
\[
c=2+\frac{|d|}{2\sqrt7},\qquad |d|\le\frac4{\sqrt7}.
\]
Increasing both diagonal entries equally to obtain the prescribed $c$
then gives
\[
\langle A_D(e)x,x\rangle\ge
\frac{c-2-|d|/(2\sqrt7)}{2(1+e)}\|x\|_{L^2}^2>0
\qquad(x\ne0),
\]
which completes the proof.
\end{proof}


\subsection{The Lagrangian discriminant}
\begin{figure}[tb]
\centering\includegraphics[width=.76\textwidth]{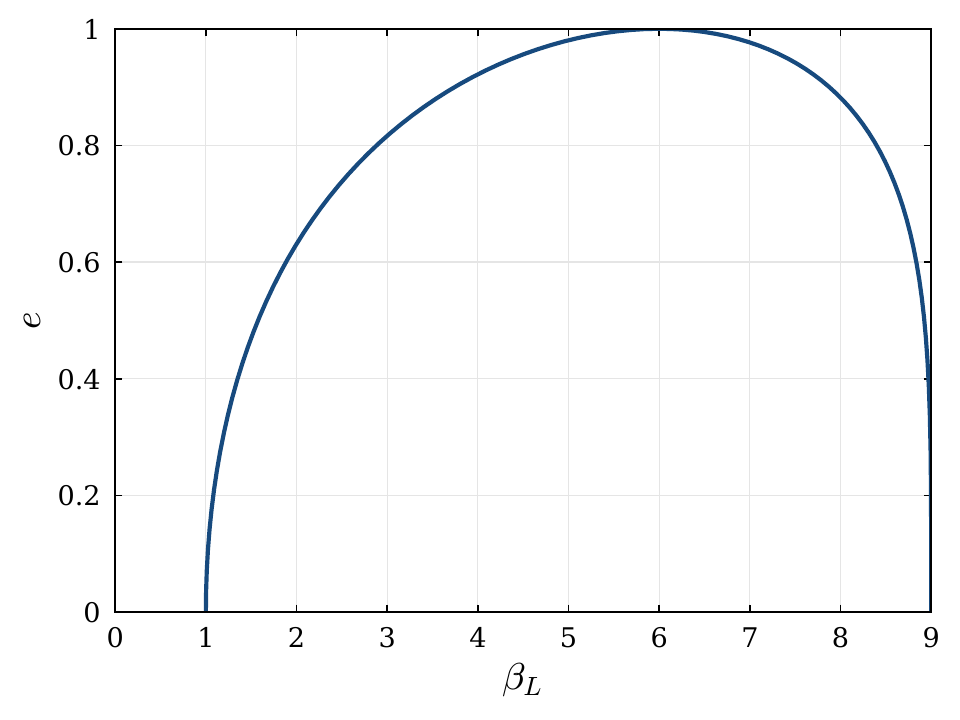}
\caption{The algebraic locus $\Delta_L=0$ in the Lagrangian parameter plane.
The labels refer to the Riccati discriminant. Their interpretation as
HH--CS interfaces. }\label{fig:discriminant}
\end{figure}

In the triangular restricted problem, the normalized primary masses are
$\mu$ and $1-\mu$. Up to interchanging the diagonal entries,
\[
c_1=\frac32(1-\sqrt{1-3\mu(1-\mu)}),\qquad
c_2=\frac32(1+\sqrt{1-3\mu(1-\mu)}).
\]
With $g=3\mu(1-\mu)$, the parameter identification is
$\beta_L=9g=27\mu(1-\mu)$, and
\[
\Delta=1-9g+2e^2+\frac{e^4}{1-g},\qquad
\Delta=0\ \Longleftrightarrow\
e^2=\sqrt{8g(1-g)}-(1-g).
\]
The unrestricted Lagrangian family has the same coefficient sum
$c_1+c_2=3$, with the larger parameter range $0\le\beta_L\le9$.
We write $\beta=\beta_L$ below.
For \eqref{eq:lagD} with $0<\beta<9$, one has $c_1+c_2=3$, $F=-\beta$,
$q=0$, and $p_1=1+e\cos t$. Formula \eqref{eq:Delta} reduces to
\begin{equation}\label{eq:DeltaL}
\Delta_L=1-\beta+2e^2+\frac{9e^4}{9-\beta}.
\end{equation}
For $0<e<1$, define
\begin{equation}\label{eq:branches}
\beta_\pm(e)=5+e^2\pm2\sqrt{2(1-e^2)(2+e^2)}.
\end{equation}
Then
\begin{equation}\label{eq:factorL}
\Delta_L=\frac{(\beta-\beta_-(e))(\beta-\beta_+(e))}{9-\beta}.
\end{equation}
Equivalently, its zero set is
\begin{equation}\label{eq:boundaryL}
e^2=\frac{\sqrt{8\beta(9-\beta)}-(9-\beta)}9.
\end{equation}
Figure~\ref{fig:discriminant} shows the two branches, which limit to
$(1,0)$ and $(9,0)$ and meet at $(6,1)$ in the closure. The point $(9,0)$ is a limit of the formula, not a point where
\eqref{eq:DeltaL} is defined. At $e=0$, the regular collision is $\beta=1$.

\begin{corollary}
In the regular Lagrangian parameter range, CS spectra can occur only between
$\beta_-(e)$ and $\beta_+(e)$. At parameters strictly between them with
$\kappa\ne0$, the spectrum is CS. Each portion of either branch satisfying
the hyperbolicity and trace conditions in \cref{thm:spectral} is an exact
local HH--CS interface.
\end{corollary}
The algebraic curves therefore supply information that complements the
unit-circle degeneracy curves: the former concern collisions of reciprocal
pairs, while the latter determine where a specific Floquet multiplier is
reached. This distinction will be important for the antiperiodic estimate.

\subsection{Euler and Robe families}
The same construction applies to other choices of $c_1,c_2$. We record
explicit formulas without extending the Lagrangian crossing theorem to
families for which its hypotheses have not been established.

At a collinear restricted libration point, with primary separation normalized
to one, put $a=(1-\mu)/r_1^3+\mu/r_2^3>1$, where $r_1,r_2$ are the
distances to the primaries. For the general Euler configuration, normalize
$m_1+m_2+m_3=1$. Its parameter is \cite{ZL2017}
\[
\beta_E=\frac{m_1(3x^2+3x+1)+m_3x^2(x^2+3x+3)}
{x^2+m_2[(x+1)^2(x^2+1)-x^2]},
\]
where $x>0$ is the unique positive root of
\begin{align*}
0={}&(m_3+m_2)x^5+(3m_3+2m_2)x^4+(3m_3+m_2)x^3\\
&-(3m_1+m_2)x^2-(3m_1+2m_2)x-(m_1+m_2).
\end{align*}
The essential diagonal matrix is $\diag(2\beta_E+3,-\beta_E)$,
with physical parameter range $[0,7]$. Thus $a=\beta_E+1$ identifies
the two coefficient families.
For $c_1=2a+1$, $c_2=1-a$,
\begin{equation}\label{eq:DeltaE}
\Delta_E=9a^2-8a+\frac{4a(a+3)e^2+4e^4}{(3a+1)^2}.
\end{equation}
The physical Euler parameter is $\beta_E=a-1\ge0$, and \eqref{eq:DeltaE}
is strictly positive there.

\begin{figure}[tb]
\centering\includegraphics[width=.76\textwidth]{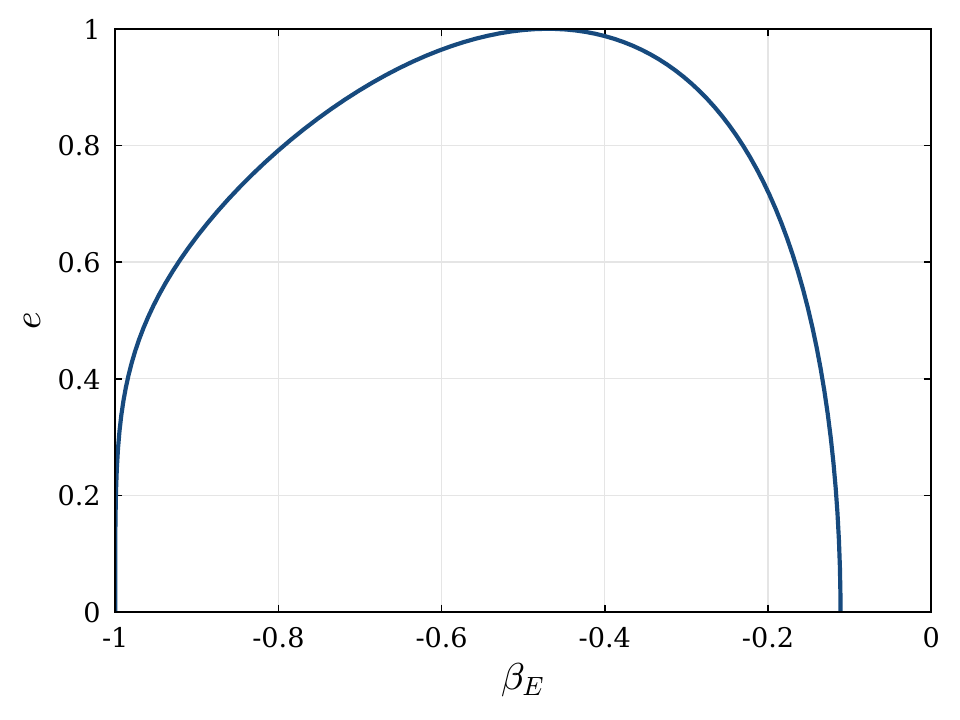}
\caption{The algebraic collision locus in the Robe extension
$-1\le\beta_E\le0$. Spectral interpretation is subject to
\cref{thm:spectral}; $e=1$ denotes a limiting value.}\label{fig:euler-robe}
\end{figure}

Robe's restricted three-body problem provides a physical realization of
the complementary parameter range. The model describes a small solid
sphere inside a fluid-filled spherical shell, subject to the attraction
of an exterior primary, and was introduced to study oscillations of the
Earth's inner core under lunar attraction. For the elliptic equilibrium
studied in \cite{ZZ2017}, the planar linearized Hamiltonian system has
the same form as the Euler essential system; the difference is that
$\beta_E\in[-1,0]$, or equivalently $a\in[0,1]$.
Thus the Euler and Robe stability problems can be treated within one
coefficient family, although their physical parameter intervals differ.
In this extension, \eqref{eq:DeltaE} can vanish, giving the algebraic
collision locus
\begin{equation}\label{eq:Robe}
e^2=\sqrt{2a(1+6a+3a^2-10a^3)}-\frac12a(a+3),
\qquad 0\le a\le\frac89.
\end{equation}
The locus is plotted in Figure~\ref{fig:euler-robe}.
Its zero-eccentricity endpoints are $\beta_E=-1$ and $\beta_E=-1/9$.
Its point at $e=1$ is
$\beta_E=(\sqrt{73}-17)/18$. For $a>0$, $q=1/(3a+1)$, so the
periodic graph condition holds for $e<1$. The value $a=0$ is obtained as
an algebraic limit because $F=0$ there.

Equivalently, the Robe locus is
\[
e^2=\sqrt{-2\beta_E(\beta_E+1)(10\beta_E^2+27\beta_E+18)}
-\frac12(\beta_E+1)(\beta_E+4).
\]

\subsection{Convex and nonconvex (2+2) configurations}
\begin{figure}[tb]
\centering\includegraphics[width=.76\textwidth]{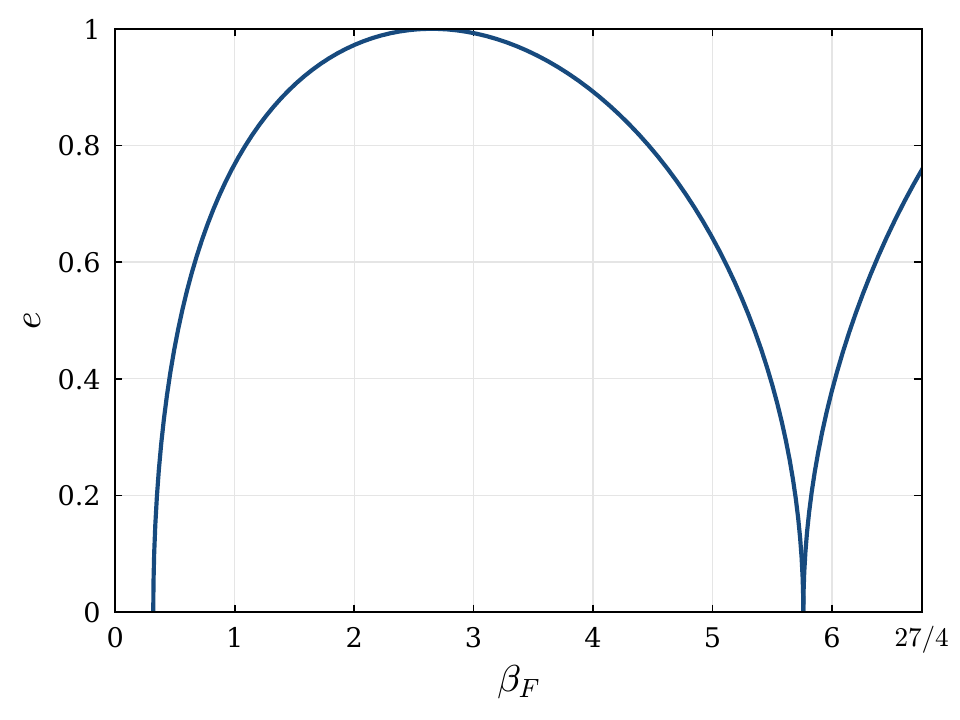}
\caption{The convex algebraic locus \eqref{eq:convexcurve} in the
$(\beta_F,e)$ plane. The point $(144/25,0)$ is an algebraic limit.}
\label{fig:convex-boundary}
\end{figure}

The mass parameter is $\beta_F=27m(1-m)$ for normalized masses
$m$ and $1-m$.
For $0\le\beta_F\le27/4$, the choices
\[
(c_1,c_2)=\left(\frac{9-3\sqrt{9-\beta_F}}{2},\sqrt{9-\beta_F}\right),\qquad
(c_1,c_2)=\left(\frac{9+3\sqrt{9-\beta_F}}{2},-\sqrt{9-\beta_F}\right)
\]
lead, respectively, to
\begin{align}
\Delta_C={}&\frac{226-25\beta_F-74\sqrt{9-\beta_F}}{4}
+\frac{12(672-75\beta_F+112\sqrt{9-\beta_F})}{(25\sqrt{9-\beta_F}-3)^2}e^2\nonumber\\
&+\frac{144(11\sqrt{9-\beta_F}+9)^2}{(5\sqrt{9-\beta_F}-9)^2(25\sqrt{9-\beta_F}-3)^2}e^4,\label{eq:DeltaC}\\
\Delta_N={}&\frac{226-25\beta_F+74\sqrt{9-\beta_F}}{4}
+\frac{12(672-75\beta_F-112\sqrt{9-\beta_F})}{(25\sqrt{9-\beta_F}+3)^2}e^2\nonumber\\
&+\frac{144(11\sqrt{9-\beta_F}-9)^2}{(5\sqrt{9-\beta_F}+9)^2(25\sqrt{9-\beta_F}+3)^2}e^4.\label{eq:DeltaN}
\end{align}
The convex zero set, away from $5\sqrt{9-\beta_F}=9$, is
\begin{equation}\label{eq:convexcurve}
\begin{aligned}
e^2={}&\frac{4|5\sqrt{9-\beta_F}-9|
\sqrt{7\sqrt{9-\beta_F}(3-\sqrt{9-\beta_F})
(\sqrt{9-\beta_F}+1)(25\sqrt{9-\beta_F}+3)^3}}
{24(11\sqrt{9-\beta_F}+9)^2}\\
&-\frac{(5\sqrt{9-\beta_F}-9)^2
(672-75\beta_F+112\sqrt{9-\beta_F})}
{24(11\sqrt{9-\beta_F}+9)^2}.
\end{aligned}
\end{equation}
Figure~\ref{fig:convex-boundary} illustrates this locus in the
$(\beta_F,e)$ plane. Only points for which the right-hand side belongs
to $[0,1)$ are regular parameters of interest. Its algebraic range starts at
$\beta_F=16(182-37\sqrt{21})/625$. The point $\beta_F=144/25$, $e=0$
is excluded from the Riccati formula because $c_1=c_2$; it is not an
additional regular zero of \eqref{eq:DeltaC}. The limiting point with $e=1$
has $\beta_F=1656/625$.
For the nonconvex choice, writing $\Delta_N=a_N+b_Ne^2+c_Ne^4$ gives
\[
\begin{aligned}
b_N^2-4a_Nc_N
&=-\frac{16128\sqrt{9-\beta_F}(\sqrt{9-\beta_F}+3)
(\sqrt{9-\beta_F}-1)}{(5\sqrt{9-\beta_F}+9)^2}
\frac{(25\sqrt{9-\beta_F}-3)^3}{(25\sqrt{9-\beta_F}+3)^4}<0,\\
c_N&>0.
\end{aligned}
\]
for $0\le\beta_F\le27/4$. 
Thus $\Delta_N>0$ throughout this range.

\section{Positive symplectic crossings and parameter monotonicity}\label{sec:positive}
The Riccati discriminant describes how reciprocal spectral pairs meet.
To determine the direction in which a unit-circle multiplier is crossed,
we now use the self-adjoint operator \eqref{eq:A}. This second description
also fixes the index convention needed for the estimates in the next section.

\subsection{A positive scalar comparison operator}
\begin{lemma}\label{lem:K}
For $0\le e<1$ and every $\omega\in\U$, the operator
\[
K_e=A(9,e)=-\frac{d^2}{dt^2}I_2-I_2+\frac{3}{2(1+e\cos t)}I_2
\]
is strictly positive on $\overline D(\omega,2\pi)$. More precisely,
\begin{equation}\label{eq:ground}
\ip{K_ex}{x}_{L^2}
=\int_0^{2\pi}\left[(1+e\cos t)^2\left|\left(\frac{x}{1+e\cos t}\right)'\right|^2
+\frac{|x|^2}{2(1+e\cos t)}\right]\dd.
\end{equation}
\end{lemma}
\begin{proof}
Since $-(1+e\cos t)''-(1+e\cos t)+1=0$, integration by parts gives
$\int(|x'|^2+(-1+1/(1+e\cos t))|x|^2)\dd
=\int (1+e\cos t)^2|(x/(1+e\cos t))'|^2\dd$.
The boundary term vanishes because $1+e\cos t$ and its derivative are periodic and $|\omega|=1$.
Adding $\int |x|^2/(2(1+e\cos t))\dd$ proves the identity and positivity.
\end{proof}

\begin{proposition}\label{prop:monotone}
For fixed $e<1$, every $\omega$-Morse index of $A(\beta,e)$ is
nonincreasing in $\beta$. At a nonzero $x\in\ker A(\beta,e)$,
\begin{equation}\label{eq:crossoperator}
\ip{\partial_\beta A(\beta,e)x}{x}_{L^2}
=\frac{\ip{K_ex}{x}_{L^2}}{2(9-\beta)}>0.
\end{equation}
\end{proposition}
\begin{proof}
For $0<\beta<9$,
\[
\frac{A(\beta,e)}{\sqrt{9-\beta}}=\frac{K_e}{\sqrt{9-\beta}}+\frac{S(t)}{2(1+e\cos t)},
\qquad
\partial_\beta\left(\frac{A}{\sqrt{9-\beta}}\right)=\frac{K_e}{2(9-\beta)^{3/2}}>0.
\]
Positive scalar normalization preserves the negative index, so the minimax
principle proves monotonicity. On the kernel, taking the inner product of
$K_ex+\sqrt{9-\beta}Sx/(2(1+e\cos t))=0$ with $x$ and using
$\partial_\beta A=-S/(4\sqrt{9-\beta}(1+e\cos t))$ proves \eqref{eq:crossoperator}.
\end{proof}

\subsection{The monodromy crossing form}
A sign convention is essential when translating \eqref{eq:crossoperator}
into a statement about symplectic paths. Put $z=(-x',x)^T$ and define
\begin{equation}\label{eq:xi}
\xi'=J_4X_{\beta,e}(t)\xi,\qquad \xi(0)=I_4,
\qquad
X_{\beta,e}=\diag\left(-I_2,-I_2+\frac{3I_2+\sqrt{9-\beta}S(t)}{2(1+e\cos t)}\right).
\end{equation}
This is precisely the same second-order equation $A(\beta,e)x=0$.
If $H=\diag(-I_2,I_2)$, then
\begin{equation}\label{eq:conventions}
\xi(t)=H\diag(R(t),R(t))\gamma_{\beta,e}(t)H,
\qquad \xi(2\pi)=H\gamma_{\beta,e}(2\pi)H.
\end{equation}
Here $H^TJ_4H=-J_4$. Thus the two monodromy matrices have identical
spectra, but the sign of their parameter generators is reversed by this
antisymplectic change of convention.

\begin{definition}
For a differentiable symplectic path $M(\beta)$, write
$\partial_\beta M=J_4\mathcal C_\beta M$, where $\mathcal C_\beta$ is real
symmetric. The path is positive in the usual sense if $\mathcal C_\beta>0$
on $\R^4$. We say that its unit-circle crossing forms are positive if
$\mathcal C_\beta$ is positive on each nonzero complex eigenspace
$\ker(M-\omega I_4)$, $\omega\in\U$, using the Hermitian extension.
\end{definition}

\begin{theorem}\label{thm:crossing}
Fix $0\le e<1$ and let $M(\beta)=\xi_{\beta,e}(2\pi)$, with \eqref{eq:xi}.
For $0<\beta<9$, the parameter generator satisfies
\begin{equation}\label{eq:generator}
\mathcal C_{\beta,e}
=\int_0^{2\pi}(\xi(t)M^{-1})^T
\partial_\beta X_{\beta,e}(t)(\xi(t)M^{-1})\dd.
\end{equation}
For every $\omega\in\U$ and nonzero $\eta\in\ker(M-\omega I_4)$,
\begin{equation}\label{eq:positiveidentity}
\eta^*\mathcal C_{\beta,e}\eta
=\frac{\ip{K_ex}{x}_{L^2}}{2(9-\beta)}>0,
\end{equation}
where $x$ is the position component of $\xi(t)\eta$.
\end{theorem}
\begin{proof}
Differentiating the Hamiltonian equation with respect to $\beta$ and using
symplecticity gives \eqref{eq:generator}. Since
$M^{-1}\eta=\omega^{-1}\eta$ and $|\omega|=1$, evaluation on $\eta$ gives
\[
\eta^*\mathcal C_{\beta,e}\eta
=\int_0^{2\pi}x(t)^*\partial_\beta
\left(\frac{3I_2+\sqrt{9-\beta}S(t)}{2(1+e\cos t)}\right)x(t)\dd
=\ip{\partial_\beta A(\beta,e)x}{x}_{L^2}.
\]
The function $x$ belongs to $\ker A(\beta,e)$ with multiplier $\omega$.
It cannot vanish identically, since the other component is $-x'$ and
$\eta\ne0$. Now apply \cref{prop:monotone} and \cref{lem:K}.
\end{proof}

This proof gives a direct operator explanation of the positive crossing
property considered in the theory of convex symplectic paths
\cite{CLW2019}. It does not assert that \eqref{eq:generator} is positive
on vectors outside the unit-circle eigenspaces. In particular, one must
not replace \cref{thm:crossing} by global positivity of the full parameter
path without an additional argument. With the convention \eqref{eq:ham},
the corresponding restricted generator is negative, as follows from
\eqref{eq:conventions}; the operator index monotonicity is unchanged.

At a fixed $\omega$-degeneracy, \eqref{eq:crossoperator} says that every
zero eigenvalue of the self-adjoint problem crosses towards positive
values as $\beta$ increases. The Morse index therefore drops by the
nullity. This gives the orientation of each degeneracy curve and explains
why a single negative test direction supplies a lower bound for the
rightmost antiperiodic curve.

\section{An estimate for the second antiperiodic degeneracy curve}\label{sec:estimate}
We now turn from the location and orientation of spectral collisions to a
quantitative estimate. Let $\beta_s(e)\le\beta_m(e)$ denote the two
antiperiodic degeneracy curves of the Lagrangian family. The established
index description \cite[Theorem 1.2]{HLS2014} is
\begin{equation}\label{eq:indexregions}
\phi_{-1}(A(\beta,e))=
\begin{cases}
2,&0\le\beta<\beta_s(e),\\
1,&\beta_s(e)\le\beta<\beta_m(e),\\
0,&\beta_m(e)\le\beta\le9.
\end{cases}
\end{equation}
When the two curves coincide, the middle interval is empty. Both start at
$(3/4,0)$ and tend to $(0,1)$. The second curve is the larger value
$\beta_m(e)$ for fixed $e$; no global monotonicity in $e$ is assumed.

\subsection{A two-frequency trial function}
Put $w=\sqrt{1-e^2}$ and let $b(e)$ be defined by \eqref{eq:b}.
This function is the unique zero in $(0,9)$ of
\begin{equation}\label{eq:G}
G(\beta,e)=8\beta-4e\sqrt{9-\beta}+(1+w)(12-15w).
\end{equation}
For $e=0$, this zero is $3/4$. For $0<e<1$, it remains positive:
if $w\ge4/5$, then $G(0,e)<0$ immediately; if $w<4/5$, the inequality
$(1+w)(4-5w)<4\sqrt{1-w^2}$ follows by squaring, since
\[
(1+w)^2(4-5w)^2-16(1-w^2)
=w(1+w)(25w^2-15w-8)<0.
\]
Also $G(9,e)>0$ and $\partial_\beta G=8+2e/\sqrt{9-\beta}>0$.

\begin{theorem}\label{thm:trial}
For $0\le e<1$ and $0\le\beta<b(e)$, there exists an explicit nonzero
$x_{\beta,e}\in\overline D(-1,2\pi)$ such that
$\ip{A(\beta,e)x_{\beta,e}}{x_{\beta,e}}_{L^2}<0$.
Consequently $\phi_{-1}(A(\beta,e))\ge1$.
For $0<e<1$, the same index inequality holds at $\beta=b(e)$.
\end{theorem}
\begin{proof}
Set $d_1=(3+\sqrt{9-\beta})/2$, $d_2=(3-\sqrt{9-\beta})/2$, and
\[
\rho=\frac e{1+w},\qquad u=\frac{1-\rho}{w},\qquad
v=\frac{1+\rho}{w},\qquad \kappa=\left(\frac14+d_2v\right)^{-1}.
\]
Choose
\begin{equation}\label{eq:trial}
x_{\beta,e}(t)=R(t)
\begin{pmatrix}\cos(t/2)\\-\kappa\sin(t/2)\end{pmatrix}.
\end{equation}
This function and its derivative are antiperiodic. Since
\[
R^TA(\beta,e)R=-\frac{d^2}{dt^2}I_2-2J_2\frac d{dt}+D_\beta/(1+e\cos t),
\]
and
\[
\int_0^{2\pi}\frac{\cos^2(t/2)}{1+e\cos t}\dd=\pi u,
\qquad
\int_0^{2\pi}\frac{\sin^2(t/2)}{1+e\cos t}\dd=\pi v,
\]
we obtain
\begin{align}
\ip{Ax_{\beta,e}}{x_{\beta,e}}_{L^2}
&=\pi\left[\frac14+d_1u-2\kappa+
\left(\frac14+d_2v\right)\kappa^2\right]\nonumber\\
&=\frac\pi{\frac14+d_2v}
\left[\left(\frac14+d_1u\right)\left(\frac14+d_2v\right)-1\right]\nonumber\\
&=\frac{\pi G(\beta,e)}{16w(1+w)(\frac14+d_2v)}.\label{eq:trialvalue}
\end{align}
For the last equality, use $uv=2/[w(1+w)]$. Since $G$ is increasing in
$\beta$, \eqref{eq:trialvalue} is negative for $\beta<b(e)$.

At $\beta=b(e)$, let $x_0$ be \eqref{eq:trial}. Its quadratic form is zero,
but $Ax_0\ne0$ for $e>0$. Indeed, if $Ax_0=0$, the function
$y=R^Tx_0$ would solve $(1+e\cos t)(y''+2J_2y')-D_\beta y=0$ and have only
frequencies $\pm1/2$. Its nonzero coefficient $a$ at frequency $1/2$
would then satisfy
\[
\frac e2\left(-\frac14I_2+iJ_2\right)a=0
\]
by comparison at frequency $3/2$. This is impossible because the matrix
has determinant $-15/16$.
The smooth antiperiodic function $x_\varepsilon=x_0-\varepsilon Ax_0$
therefore satisfies
\[
\ip{Ax_\varepsilon}{x_\varepsilon}_{L^2}
=-2\varepsilon\|Ax_0\|_{L^2}^2
+\varepsilon^2\ip{A(Ax_0)}{Ax_0}_{L^2}<0
\]
for sufficiently small $\varepsilon>0$. This proves the endpoint claim.
\end{proof}

\begin{corollary}\label{cor:second}
For $0<e<1$, the second antiperiodic degeneracy curve satisfies
\begin{equation}\label{eq:strictbound}
\beta_m(e)>b(e).
\end{equation}
Moreover,
\[
b(0)=\frac34,\qquad b'(0)=\frac{\sqrt{33}}4,
\qquad \lim_{e\to1^-}b(e)=0.
\]
Thus the explicit bound has the same initial value and tangent as
$\beta_m$ at the circular limit.
\end{corollary}
\begin{proof}
At $\beta=b(e)$ the index is at least one by \cref{thm:trial}, whereas
\eqref{eq:indexregions} gives index zero for $\beta\ge\beta_m(e)$.
This proves the strict inequality. The endpoint values follow from
\eqref{eq:b}. Differentiation of $G(b(e),e)=0$ at $e=0$ gives
$b'(0)=\sqrt{33}/4$, which is the known initial slope of the second
curve \cite[Proposition 7.4]{HLS2014}.
\end{proof}

\begin{figure}[tb]
\centering\includegraphics[width=.75\textwidth]{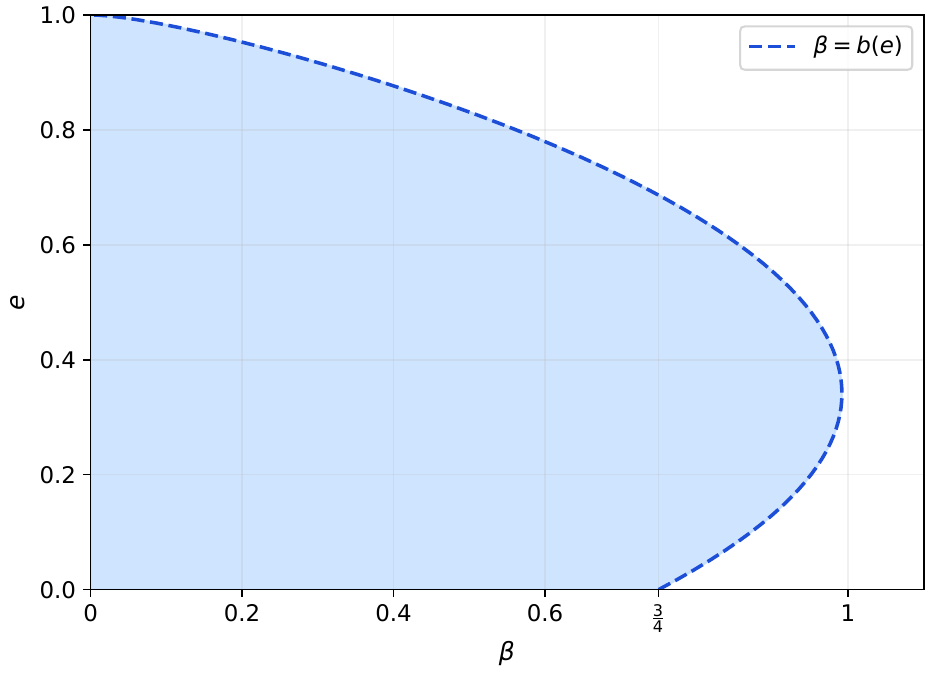}
\caption{The explicit trial-function boundary $\beta=b(e)$. The shaded
region has antiperiodic Morse index at least one. For $0<e<1$, the second
$-1$-degeneracy curve lies strictly to the right of the displayed curve.
The line itself is not asserted to be a degeneracy curve.}\label{fig:trial}
\end{figure}

\subsection{Comparison bounds and their limitations}
The preceding lower bound for $\beta_m$ can be placed in a larger enclosure
using constant-denominator comparison operators. For $\eta>0$, put
\[
B_\eta(\beta)=-\frac{d^2}{dt^2}I_2-I_2+
\frac{3I_2+\sqrt{9-\beta}S(t)}{2\eta}.
\]
The matrix numerator is positive semidefinite, so
\begin{equation}\label{eq:comparison}
B_{1+e}(\beta)\le A(\beta,e)\le B_{1-e}(\beta).
\end{equation}
After rotation, the Fourier matrix at $k\in\mathbb Z+1/2$ has positive
trace and determinant
\[
k^4+\left(\frac3\eta-4\right)k^2+\frac\beta{4\eta^2}.
\]
For $0<\eta<2$ and $\beta\ge0$, only $|k|=1/2,3/2$ can give a
nonnegative degeneracy parameter. The corresponding values are
\[
f_1(\eta)=\frac{15}4\eta^2-3\eta,
\qquad f_3(\eta)=\frac{63}4\eta^2-27\eta,
\]
and $f_1-f_3=12\eta(2-\eta)>0$. Hence $B_\eta\ge0$ when
$\beta\ge f_1(\eta)$, while $\beta<f_1(\eta)$ gives two negative
Fourier directions at $k=\pm1/2$. Combining this fact with
\eqref{eq:comparison} and \eqref{eq:indexregions} yields
\begin{equation}\label{eq:enclosure}
\max\left\{0,\frac34-\frac92e+\frac{15}4e^2\right\}
\le\beta_s(e)\le\beta_m(e)
\le\frac34+\frac92e+\frac{15}4e^2.
\end{equation}
The leftmost estimate is nontrivial only for $e<1/5$.
The additional estimate \eqref{eq:strictbound} concerns the second curve
specifically and is tangent to it at $e=0$.

These results describe distinct features of one stability diagram.
The algebraic locus \eqref{eq:boundaryL} concerns reciprocal-pair collisions,
\cref{thm:crossing} fixes the orientation of unit-circle crossings, and
\eqref{eq:strictbound} supplies a computable lower bound for the second
antiperiodic curve. None of these statements requires identifying a trial
function of zero quadratic form with an exact zero eigenfunction.

\appendix
\section{Scalar identities underlying the diffential Riccati equation}\label{app:riccati}
\subsection{The graph transformation and scalar equations}
The block identity behind the graph transformation is
\[
J_4B_DT=
\begin{pmatrix}
-J_2Q_+-I_2+D/(1+e\cos t)&-J_2Q_--I_2+D/(1+e\cos t)\\
Q_+-J_2&Q_--J_2
\end{pmatrix},
\]
whereas
\[
T'+T\diag(P_+,P_-)=
\begin{pmatrix}Q_+'+Q_+P_+&Q_-'+Q_-P_-\\P_+&P_-\end{pmatrix}.
\]
Equality of the lower and upper rows gives $P_\pm=Q_\pm-J_2$ and
\eqref{eq:riccati}, respectively.

\subsection{Symmetric and skew-symmetric parts}
We construct particular solutions of \eqref{eq:riccati} for the constant
diagonal matrix $D=\diag(c_1,c_2)$. Write $Q=(Q_{ij})$ and set
\[
S=\frac{Q+Q^T}{2},\qquad \Omega=\frac{Q-Q^T}{2},\qquad Q=S+\Omega.
\]
Here $S$ denotes the symmetric part of $Q$, rather than the rotating
matrix $S(t)$ in \eqref{eq:A}. Taking the symmetric and skew-symmetric
parts of the Riccati equation yields
\begin{align}
S'&=-S^2-\Omega^2+SJ_2-J_2S-I_2+\frac{D}{1+e\cos t},
\label{app:symmetric}\\
\Omega'&=-S\Omega-\Omega S+\Omega J_2-J_2\Omega.
\label{app:skew}
\end{align}
Every skew-symmetric $2\times2$ matrix has the form $\Omega=\omega J_2$.
Since $SJ_2+J_2S=(\tr S)J_2$, equation \eqref{app:skew} becomes
\[
\omega'J_2=-\omega(\tr S)J_2,
\]
and hence
\begin{equation}\label{app:omega}
\omega'=-\omega\tr S.
\end{equation}
The branch $\omega\equiv0$ consists of symmetric solutions. On a connected
interval where $Q$ is defined, any other branch has $\omega\ne0$
throughout, by uniqueness for \eqref{app:omega}. On such an interval, put
\begin{equation}\label{app:fg}
f=\frac1\omega=\frac2{Q_{21}-Q_{12}},\qquad
 g=\frac f2(Q_{12}+Q_{21}).
\end{equation}
These unnormalized functions will give the functions $p_1,p_2$ used in
the main text after division by a constant. Equation \eqref{app:omega}
gives
\begin{equation}\label{app:trace}
Q_{11}+Q_{22}=\tr S=\frac{f'}f.
\end{equation}
The off-diagonal entry of \eqref{app:symmetric} gives
\begin{align*}
(Q_{12}+Q_{21})'&=2S_{12}'\\
&=-2S_{12}(S_{11}+S_{22})+2(S_{22}-S_{11})\\
&=-(Q_{12}+Q_{21})(Q_{11}+Q_{22})+2(Q_{22}-Q_{11}).
\end{align*}
Differentiating the second definition in \eqref{app:fg}, we therefore obtain
\begin{align}
g'&=\frac{f'}2(Q_{12}+Q_{21})+\frac f2(Q_{12}+Q_{21})'\nonumber\\
&=\frac f2(Q_{11}+Q_{22})(Q_{12}+Q_{21})\nonumber\\
&\quad+\frac f2\bigl[-(Q_{12}+Q_{21})(Q_{11}+Q_{22})
+2(Q_{22}-Q_{11})\bigr]\nonumber\\
&=f(Q_{22}-Q_{11}).\label{app:gprime}
\end{align}
Consequently, all four entries are recovered as
\[
Q_{11}=\frac{f'-g'}{2f},\quad Q_{22}=\frac{f'+g'}{2f},\qquad
Q_{12}=\frac{g-1}{f},\quad Q_{21}=\frac{g+1}{f}.
\]
In matrix form,
\begin{align}
Q&=\frac1f\begin{pmatrix}(f'-g')/2&g-1\\g+1&(f'+g')/2\end{pmatrix},
\label{app:Qfg}\\
S&=\frac1f\begin{pmatrix}(f'-g')/2&g\\g&(f'+g')/2\end{pmatrix},
\qquad \Omega=\frac1fJ_2.\nonumber
\end{align}

\subsection{Reduction to two scalar equations}
Subtracting the diagonal entries of \eqref{eq:riccati} gives
\begin{align}
(Q_{22}-Q_{11})'
={}&-(Q_{11}+Q_{22})(Q_{22}-Q_{11})\nonumber\\
&-2(Q_{12}+Q_{21})+\frac{c_2-c_1}{1+e\cos t}.
\label{app:diagonal-difference}
\end{align}
Using \eqref{app:trace} and \eqref{app:gprime}, we find
\begin{align*}
g''&=f'(Q_{22}-Q_{11})+f(Q_{22}-Q_{11})'\\
&=f(Q_{11}+Q_{22})(Q_{22}-Q_{11})\\
&\quad-f(Q_{11}+Q_{22})(Q_{22}-Q_{11})
-2f(Q_{12}+Q_{21})+\frac{c_2-c_1}{1+e\cos t}f\\
&=-4g+\frac{c_2-c_1}{1+e\cos t}f.
\end{align*}
Thus the first scalar equation is
\begin{equation}\label{app:scalar-g}
g''+4g=\frac{c_2-c_1}{1+e\cos t}f.
\end{equation}
On the other hand, taking the trace of \eqref{eq:riccati} yields
\begin{equation}\label{app:trace-equation}
(\tr Q)'=-\tr(Q^2)-2+\frac{c_1+c_2}{1+e\cos t}.
\end{equation}
Formula \eqref{app:Qfg} gives explicitly
\[
\tr(Q^2)=\frac{(f')^2+(g')^2}{2f^2}
+\frac{2(g^2-1)}{f^2},\qquad
(\tr Q)'=\frac{ff''-(f')^2}{f^2}.
\]
Substituting these expressions into \eqref{app:trace-equation} and
multiplying by $f^2$ gives
\[
ff''-(f')^2=-\frac{(f')^2+(g')^2}{2}-2g^2+2
-\left(2-\frac{c_1+c_2}{1+e\cos t}\right)f^2.
\]
Equivalently, the second scalar equation is
\begin{equation}\label{app:scalar-f}
ff''-\frac12(f')^2+\left(2-\frac{c_1+c_2}{1+e\cos t}\right)f^2
+2g^2+\frac12(g')^2=2.
\end{equation}
Conversely, if $f\ne0$ and \eqref{app:scalar-g}--\eqref{app:scalar-f}
hold, substitution of \eqref{app:Qfg} recovers all entries of
\eqref{eq:riccati}. The reduction is therefore equivalent on this branch.

\subsection{Trigonometric coefficient comparison}
Take
\[
f=(1+e\cos t)(k_0+k_1\cos t),\qquad
 g=\frac{(c_2-c_1)k_0}{4}
 +\frac{(c_2-c_1)k_1}{3}\cos t+k_2\cos2t.
\]
With this ansatz for $f$, equation \eqref{app:scalar-g} reads
\[
g''+4g=(c_2-c_1)(k_0+k_1\cos t).
\]
The displayed expression for $g$ solves this equation. We choose the even solution of the
forced equation for $g$; its general solution could also contain a
$\sin2t$ term. Substitution in the second scalar identity yields
\[
A_0+A_1\cos t+A_2\cos2t+A_3\cos3t=0,
\]
where
\begin{align}
A_0={}&
\left[\frac{(c_2-c_1)^2}{8}+\frac{e^2}{4}+2-c_1-c_2\right]k_0^2+\left[\frac{5(c_2-c_1)^2}{36}+\frac14-\frac{c_1+c_2}{2}\right]k_1^2\nonumber\\
&\quad+e\left(\frac52-c_1-c_2\right)k_0k_1+2k_2^2-2,
\label{app:A0}\\[4pt]
A_1={}&
\left[\frac{(c_2-c_1)^2}{3}+\frac{3e^2}{4}+3-2(c_1+c_2)\right]k_0k_1\nonumber\\
&+(c_2-c_1)k_1k_2
+e(3-c_1-c_2)k_0^2
+\frac{3e}{4}(1-c_1-c_2)k_1^2,
\label{app:A1}\\[4pt]
A_2={}&
\left[\frac{(c_2-c_1)^2}{12}+\frac34-\frac{c_1+c_2}{2}\right]k_1^2+(c_2-c_1)k_0k_2+\frac{3e^2}{4}k_0^2
+e\left(\frac32-c_1-c_2\right)k_0k_1,
\label{app:A2}\\[4pt]
A_3={}&
\frac{k_1}{12}
\left[4(c_2-c_1)k_2+3e^2k_0
-3e(c_1+c_2-1)k_1\right].  \label{app:A3}
\end{align}

Assume $c_1\ne c_2$ and $F\ne0$, as in \eqref{eq:nondeg}.
For $k_1\ne0$, $A_3=0$ gives
\[
k_2=\frac{3e}{4(c_2-c_1)}[(c_1+c_2-1)k_1-ek_0].
\]
With this substitution, $A_2=0$ becomes
\[
\frac{k_1}{12}[Fk_1-3e(c_1+c_2-3)k_0]=0.
\]
Since $F=(c_2-c_1)^2-6(c_1+c_2)+9$, this yields
\[
k_1=\frac{3e(c_1+c_2-3)}{(c_2-c_1)^2-6(c_1+c_2)+9}k_0.
\]
Substitution into the formula for $k_2$ then gives
\[
k_2=\frac{3e^2[2c_1^2+8c_1c_2+2c_2^2-6(c_1+c_2)]}
{4(c_2-c_1)[(c_2-c_1)^2-6(c_1+c_2)+9]}k_0.
\]
In terms of $q$ from \eqref{eq:nondeg}, the regular branch is
\[
k_1=eqk_0,\qquad
k_2=-\frac{3e^2}{4(c_2-c_1)}[1-(c_1+c_2-1)q]k_0.
\]
These expressions make $A_1,A_2,A_3$ vanish identically, including
when $k_1=0$; hence no division by $k_1$ is needed to verify those cases.
The last equation is
\[
A_0=\frac{\Delta}{8}k_0^2-2=0,\qquad
k_0=\pm\frac4{\sqrt\Delta}\quad(\Delta\ne0).
\]
The coefficient $\Delta$ is exactly \eqref{eq:Delta}. To display its
polynomial structure without an additional abbreviation, it can also be written as
\begin{align*}
\Delta={}&(c_2-c_1)^2-8(c_1+c_2-2)\\
&+\frac{4e^2}{[(c_1+c_2-3)^2-4c_1c_2]^2}
\Bigl[-3(c_1+c_2-3)^4+3(c_1+c_2-3)^3\\
&\hspace{28mm}+10c_1c_2(c_1+c_2-3)^2
+12c_1c_2(c_1+c_2-3)+8c_1^2c_2^2\Bigr]\\
&+\frac{9e^4}{(c_2-c_1)^2}
\left[1-\frac{3(c_1+c_2-3)(c_1+c_2-1)}
{(c_1+c_2-3)^2-4c_1c_2}\right]^2.
\end{align*}
For $\Delta>0$, the two choices of $k_0$ are real. For $\Delta<0$,
the construction is interpreted over $\C$, with either square root of
$\Delta$.

\subsection{Normalization and reconstruction}
Set $p_1=f/k_0$ and $p_2=g/k_0$. These functions do not depend on
the choice of sign of $k_0$ and are precisely
\begin{align*}
p_1(t)&=(1+e\cos t)(1+eq\cos t),\\
p_2(t)&=\frac{c_2-c_1}{4}+\frac{(c_2-c_1)eq}{3}\cos t
-\frac{3e^2[1-(c_1+c_2-1)q]}{4(c_2-c_1)}\cos2t.
\end{align*}
Dividing the two scalar equations by $k_0$ and $k_0^2$, respectively,
gives \eqref{eq:scalar1}--\eqref{eq:scalar2}, since $2/k_0^2=\Delta/8$.
Substitution into \eqref{app:Qfg} yields
\begin{align*}
Q_\pm
&=\frac1{p_1}
\left[\frac12(p_1'I_2-p_2'N)
+p_2\begin{pmatrix}0&1\\1&0\end{pmatrix}
\pm\frac{\sqrt\Delta}{4}J_2\right]\\
&=\frac1{p_1}
\begin{pmatrix}(p_1'-p_2')/2&p_2\mp\sqrt\Delta/4\\
p_2\pm\sqrt\Delta/4&(p_1'+p_2')/2\end{pmatrix},
\qquad N=\diag(1,-1).
\end{align*}
In particular,
\[
Q_+-Q_-=\frac{\sqrt\Delta}{2p_1}J_2,\qquad
\det T=\frac{\Delta}{4p_1^2},\qquad P_\pm=Q_\pm-J_2.
\]
For $0\le e<1$ and $|eq|<1$, $p_1$ is positive and periodic.
Thus $\Delta\ne0$ gives two complementary periodic graphs, real when
$\Delta>0$ and complex conjugate when $\Delta<0$.
When $\Delta=0$, the normalized formulas remain meaningful wherever
$p_1\ne0$ and solve the Riccati equation by direct substitution in
\eqref{eq:scalar1}--\eqref{eq:scalar2}. However, the two graphs coincide
and $T$ is singular. This case is not obtained by dividing by
$\sqrt\Delta$. The construction supplies particular solutions; it does
not classify all Riccati solutions, including symmetric solutions or
branches at the excluded coefficient values.

\subsection{The algebraic endpoint identity}
One useful algebraic endpoint check is
\[
\Delta\big|_{e=1}=
\frac{[(c_1-c_2)^2-(c_1+c_2)]^2
[(c_1-c_2)^2-9(c_1+c_2)+18]^2}
{(c_1-c_2)^2[(c_1-c_2)^2-6(c_1+c_2)+9]^2},
\]
whenever the denominator is nonzero. Since $\Delta$ depends on $e^2$,
the same identity holds at $e=-1$. Consequently, under this denominator
condition,
\[
\Delta\big|_{e=\pm1}=0
\quad\Longleftrightarrow\quad
(c_1-c_2)^2=c_1+c_2
\quad\text{or}\quad
(c_1-c_2)^2=9(c_1+c_2)-18.
\]
This identity concerns the algebraic
formula only and does not remove the singularity of the differential
system at $e=1$.


\begin{thebibliography}{99}
\small
\setlength{\itemsep}{3pt}
\bibitem{CLW2019} Y. Chang, Y. Long, and J. Wang,
On bifurcation of eigenvalues along convex symplectic paths,
\emph{Ann. Inst. H. Poincar\'e C Anal. Non Lin\'eaire} \textbf{36} (2019), 75--102.
\bibitem{HLS2014} X. Hu, Y. Long, and S. Sun,
Linear stability of elliptic Lagrangian solutions of the planar three-body
problem via index theory,
\emph{Arch. Ration. Mech. Anal.} \textbf{213} (2014), 993--1045.
\bibitem{HLO2020} X. Hu, Y. Long, and Y. Ou,
Linear stability of the elliptic relative equilibrium with $(1+n)$-gon
central configurations in planar $N$-body problem,
\emph{Nonlinearity} \textbf{33} (2020), 1016--1045.
\bibitem{MSS2006a} R. Mart\'{i}nez, A. Sam\`a, and C. Sim\'o,
Stability diagram for $4$D linear periodic systems with applications to
homographic solutions,
\emph{J. Differential Equations} \textbf{226} (2006), 619--651.
\bibitem{MSS2006b} R. Mart\'{i}nez, A. Sam\`a, and C. Sim\'o,
Analysis of the stability of a family of singular-limit linear periodic
systems in $\R^4$. Applications,
\emph{J. Differential Equations} \textbf{226} (2006), 652--686.
\bibitem{MS2005} K. R. Meyer and D. S. Schmidt,
Elliptic relative equilibria in the $N$-body problem,
\emph{J. Differential Equations} \textbf{214} (2005), 256--298.
\bibitem{Roberts2002} G. E. Roberts,
Linear stability of the elliptic Lagrangian triangle solutions in the
three-body problem,
\emph{J. Differential Equations} \textbf{182} (2002), 191--218.
\bibitem{Tschauner1973} J. Tschauner,
Das elliptische eingeschr\"ankte Dreik\"orperproblem,
\emph{Celestial Mechanics} \textbf{8} (1973), 181--187.
\bibitem{ZZ2017} Q. Zhou and Y. Zhang,
Analytic results for the linear stability of the equilibrium point in
Robe's restricted elliptic three-body problem,
\emph{Discrete Contin. Dyn. Syst.} \textbf{37} (2017), 1763--1787.
\bibitem{HO2016} X. Hu and Y. Ou,
Collision index and stability of elliptic relative equilibria in planar
$n$-body problem, \emph{Comm. Math. Phys.} \textbf{348} (2016), 803--845.
\bibitem{HS2009} X. Hu and S. Sun,
Index and stability of symmetric periodic orbits in Hamiltonian systems
with application to figure-eight orbit,
\emph{Comm. Math. Phys.} \textbf{290} (2009), 737--777.
\bibitem{HS2010} X. Hu and S. Sun,
Morse index and stability of elliptic Lagrangian solutions in the planar
three-body problem, \emph{Adv. Math.} \textbf{223} (2010), 98--119.
\bibitem{HWZ2026} X. Hu, L. Wu, and A. Zhu,
Strict separation of antiperiodic degeneracy curves for elliptic Euler
and Lagrangian solutions, preprint, arXiv:2611.01719v1 (2026).
\bibitem{Long2002} Y. Long,
\emph{Index Theory for Symplectic Paths with Applications},
Progress in Mathematics, vol. 207, Birkh\"auser, Basel, 2002.
\bibitem{RS1993} J. Robbin and D. Salamon,
The Maslov index for paths, \emph{Topology} \textbf{32} (1993), 827--844.
\bibitem{ZL2017} Q. Zhou and Y. Long,
Maslov-type indices and linear stability of elliptic Euler solutions of
the three-body problem,
\emph{Arch. Ration. Mech. Anal.} \textbf{226} (2017), 1249--1301.
\end{thebibliography}
\end{document}